\documentstyle[aps,12pt]{revtex}

\begin{document}           \title{ Theory Summary and
Future Directions \\ }

\author{David J. Gross \thanks{ Supported in part by
the National Science Foundation under \eject Grant PHY90-
21984.}\\}
\address{Physics Department\\ Princeton
University\\ Princeton, New Jersey, 08544\\}

\maketitle

\section*{Introduction}

        Since I am the last speaker, I would like, on
behalf of all the participants, speakers and listeners,
to thank the organizers of this extremely well
organized, well run and stimulating conference at
such a beautiful location. I would also like to thank
all the other speakers for giving such marvelous
talks; especially the theorists, who have given
wonderful summaries of the state of particle theory,
leaving me     to summarize the
summaries and review the reviews.  The only thing
that is really left for me to do is to tell the jokes and
provide what our ex-president called \lq \lq the
vision thing".  Unfortunately, given the clouds that
hang over our field due to the short-sightedness of
the U.S. House of Representatives, it is difficult to
produce either jokes or vision. But I'll try. What I
propose to do is to discuss what some of the
highlights and lessons of this conference, mostly those
that have to do with theoretical developments,
problems and prospects.\footnote{ I shall follow the standard practice of
refering all references \eject to the preceeding talks.}

        The main message from this meeting,  as well
as all other meetings in the last few years, is that the
standard model---the electroweak theory and QCD---works
extremely well.  We are now testing it to
unprecedented accuracy as well as deepening our
understanding of electron-electron, electron-proton,
proton-proton, scattering and exploring the spectrum
and the decays of hadronic states.

The success of the standard model  is good news.
You hear often hear  people complaining at meetings
like this that the standard model
works too well. This agreement of the theory with experiment makes many
physicists unhappy.
This is   understandable. Experimentalists
want to discover new phenomena and  new laws of
physics and besides they want to prove that the
theorists are wrong. Theorists, of course, want to get
clues as to what  the new phenomena and  the new
laws are and besides  they want to prove that the
other theorists are wrong.  But we should stop
complaining about the success of the standard model, because after all
the purpose of all of our hard work is to discover the
phenomena of nature---what goes on, and then to
understand the laws of nature---how it works and
why it works.  We should be proud and happy that we
have a theory that is very successful at what we now
refer to as  low energy physics (less than a Tev) and  that
it provides a coherent picture of the particles and their interactions that we
can understand.
That is,
after all, what our effort is all about. We should not
be worried that there will not be new physics, we
know that there will be new physics. The
inadequacies of the standard model and many other
clues  give us indication that there will be new physics at higher energy and
most
likely not too far away. If  our
representatives give us the means to look for the new physics then we
will find it.  The very success and power of the
standard model is basic and necessary if we are to succeed in finding the new
physics and make fundamental new discoveries.  You must  know
a lot if you hope to discover something profoundly
new.

         By now, given the enormous success of
the standard model, we should come up with a new
name for this theory. The  {\em \lq\lq standard model"}  is not a good
name for what is after all  a comprehensive theory of low
energy physics.   The name, standard model, was
invented, in the early seventies.  I think it was first
used by Steve Weinberg to refer to the $SU(2) \times  U(1)$
electroweak theory and to differentiate it from its
few rivals, based on the $SU(3)$ and $SU(2)_R \times SU(2)_L $
gauge groups,  alternatives that were still viable at
that time.  Later  in the middle seventies, with the
emergence of QCD, it became the name   for the combination of the
electroweak  theory and quantum chromodynamics.
However, the situation has changed.  There are no
alternatives to either $SU(2)_R \times SU(2)_L $ or to QCD.
The standard model works and is well
tested. It is clear by now that it is not a model in any
case; it is a full fledged theory, which is being tested, in many cases,
with an accuracy of better than a tenth of a
percent, with sensitivity to   higher loop effects.

        We need a new name that captures the
dignity of this theory.  It is very hard to
come up with a name and I have not done very well.  I
have a few suggestions: {\em Quantum Particle Theory}  or perhaps
QPT (the name must have a good acronym); or QPD,
{\em Quantum Particle Dynamics}, for the combination of quantum
chromodynamics and quantum flavor dynamics.
Another suggestion, in honor of one of the   missing
ingredients of the standard model, the top quark, is TOP---namely the {\em
Theory of
Particles}.  Michael Dine has a nice suggestion---TONE, for the {\em The Theory
of Nearly
Everything}.   These are only tentative suggestions.
Perhaps we should have a contest to come up with a
replacement for the standard model.  Meanwhile, I will
continue to use the standard name.

        Another joke.  This came up during lunch
conversations about the gloomy outlook for the
SSC.  Someone had the idea that we
should look for private funds to support the SSC, just as astronomers convince
  private individuals to finance large telescopes
at a level of a hundred million dollars.  But how to
attract such money?  The idea was that in order to finance the
SSC  we could auction the quarks.  We have given the
quarks peculiar names -- up, down, strange,
charm, top and bottom. However we  could now sell the names
of the quarks for large contributions to the SSC. If
you had enough money  you might be willing to spend
a billion dollars to go down  in history as a quark. We
might raise a few billion dollars that way.  The only
problem with this  idea is that   instead of reporting
that CLEO has observed  the   radiative transition of
the bottom quark  to a strange quark, we would have
to say that they've seen the radiative decay of the
David Rockefeller quark to the Ross Perot quark.
Maybe it is not worth a few billion dollars!

\section*{Electroweak Interactions}

The electroweak theory, as
  has been widely reported  at this conference, passes all tests
with increasing and impressive accuracy.  The accuracy
is now often  at the  level of  .1\% to1\%,
most importantly they are beginning to be
sensitive to higher order radiative corrections.  This is quite a remarkable
achievement; it is
staggering if one compares the situation with what it
was twenty,  or even ten years ago. One of the
examples of the confrontation between experiment
and theory  is measurement of the weak angle and the $\rho$-parameter:
\begin{eqnarray}
 \sin^2(\theta_{\rm lept.})& =& .2321 \pm .0006 \, \, [ {\rm EXP.}]\, ; \qquad
= \, \,\, .2332 \pm .0013 \, \, [ {\rm TH.}] \cr
\rho_{\rm lept.} &=& \!\!\!  1.0035 \pm .0036 \, \, [ {\rm EXP.}]\, ; \qquad =
1.0026 \pm .0038\, \, [ {\rm TH.}]
\end{eqnarray}
Many more examples were  given in  talks at this meeting.

        At this level of comparison with theory, the precision
tests of the standard electroweak theory can be used
for many purposes.  We can
use the fact that the measurements agree with the calculated  radiative
corrections to constrain the existence of
particles that we cannot observe directly but can rule
out indirectly.  One famous example is the limitation
on the number of light neutrinos.   In addition there are strong
limitations on  the masses and couplings of  extra  $U(1)$ gauge bosons
($Z$'s), and on  supersymmetric
particles.  Most amusingly, these tests can be regarded as providing indirect
measurements of the top quark mass. The preferred value is $164 \pm 17\, {\rm
Gev}$
and the range is rapidly   narrowing.   Given that the precision tests are
becoming so
accurate and the search for the top quark mass is
proceeding so slowly, it seems that we will know the
mass of the top quark to better than   ten percent
before we  discover it directly. This is quite a remarkable
achievement, especially when we note that we  have absolutely no
theoretical understanding as to the origin of the
magnitude of quark masses.

In addition, the  precision tests of the electroweak
theory are beginning to be  important tools in
ruling out or confirming various theoretical ideas
as to the origin of the electroweak symmetry
breaking or  about non-standard model physics.  It
is fair to say that technicolor  has
been killed by these experiments.  This is a nice
example of a moderately appealing theoretical idea
that can be and has been ruled out by experiment.
On the other hand, the predictions  of the minimal standard model
or of the   minimal supersymmetric  extensions of the  standard
model  are quite consistent with the
precision tests.  Most important, these precision
tests are beginning to provide one with a very solid
spring board towards extrapolation, towards
unification at very high energies. In particular
 the famous coupling
constant unification at energies near the Planck scale   gives  comfort to
believers
in supersymmetry and in string theory.

\section*{Quantum Chromodynamics}

We now turn to QCD--the other component of the
standard model. We have heard a lot about tests of
QCD in this meeting as well as measurements of properties
of hadrons of various types.  My impression of all of
this is that there has been remarkable progress in
the last few years in testing QCD in various
processes-- deep inelastic scattering, jet
production, etc.  The confrontation of experiment
with theory   has been rendered more meaningful by
high order calculations.  I agree completely with
Frank Wilczek's conclusion-QCD is right.  In
addition to testing QCD, there are many new
measurements of hadron physics.  Most impressive at
this meeting was the enormous amount of data on
heavy quark spectroscopy and heavy quark decays.
The experimental measurements are impressive  and
so are the theoretical approaches to this subject. The newly  developed
heavy quark effective theories have proven to be extremely useful in
understanding heavy quark spectroscopy. Lattice gauge
theory  has developed greatly in the last few years, not just
because of the great increase in the power of modern
computers, but also because lattice theorists are
beginning to ask new and interesting questions and
develop  more powerful calculational methods.
Finally   there has been a lot of progress in using QCD
as a theoretical tool for background calculations to
ongoing experiments at high energies and for
planning new experiments at the new accelerators.
Other uses of the theory are  for nuclear physics and
for nuclear matter  under unusual
circumstances; and as a basis of attempts to construct cosmological scenarios,
for example the
attempt  to understand the origin of baryon
number.

        I want to make one remark about tests of  QCD
which has to do with a sum rule to which I am
particularly attached  to.  When I first visited CERN
in 1969, I collaborated with a young postdoc there--
we were all young in those days--who has since gone
on to  an administrative position at CERN.
The (Gross-Llewyllyn-Smith) sum rule, which
we derived using free field current commutators, or
equivalently using  the parton model,  was
interesting because it provided a direct test of
whether the constituents of the proton had baryon
number equal to one-third.
It states that

\begin{equation}
{1 \over 2} \int dx F_3^{ \nu p + \bar \nu p}(q^2,x ) = 3 .
\end{equation}

At the time Perkins  and his group were doing
neutrino experiments at CERN and  were very
interested in checking this sum rule. They concluded
from their data that it  was correct with an accuracy
of about 100\% (a factor of 2).  That was good enough
for us to believe immediately in its validity and was
one of the reasons that convinced me that quarks
really existed inside protons.  It is very enjoyable to
come to a this conference  24 years later and to hear
the sum rule  is now being tested to
an accuracy of $\sim  2\%$. As Frank noted, the
moments of $F_3$ yield some of the best tests of the
scaling violations predicted by asymptotically free
gauge theories, and we have seen that  these  are now
tested to  the level of 2\%. But the value of the sum
rule  does not come out to be $3$ but rather $2.50 \pm 0.018 \pm 0.078$. This
was
compared in this conference with the predictions of
QCD, which of course gives corrections to our sum
rule of order $\alpha_s(q^2)$; in one loop the
corrections reduce the sum rule to be equal to $2.66\pm 0.04$.  It was
remarked that this represented a slight discrepancy
between theory and experiment.
I would like to  to correct this impression.  QCD is
not just one loop. Remarkably, the   calculation of the
sum rule has now been carried out to three loop order, with the
result that,

\begin{eqnarray}
{1\over 2}  \int dx F_3^{ \nu p + \bar \nu p}(q^2,x )& =& 3[1 - {\alpha_s \over
\pi} -{43\over 12}({\alpha_s \over \pi})^2 -18.9757 ({\alpha_s \over \pi})^3
+\dots]
\cr & \sim & 2.55\pm 0.05
\end{eqnarray}
The answer again is in
agreement with  experiment, with  an accuracy of $\sim  2\%$.

        Some of the other outstanding  problems withthe comparisons of QCD
with experiment, those having to do with the
 the EMC  effect and the validity of the
Bjorken sum rule,  seem to be on the verge of being
resolved.

Let me say a word about the theoretical situation in
QCD.  All progress
that has been  made in QCD relies on the existence of small
parameters. To paraphrase Maurice Chevalier: thank
heaven for small parameters.  In fact, most of our
understanding and comprehension of
physics relies on the existence of small parameters. The   most important small
parameters in physics, that  determine the structure of the macroscopic world
are :

a. $ {M_W\over M_{\rm Planck}}\sim 10^{-16}$.

b.  $ {\Lambda \over M^4_{\rm Planck}}\leq  10^{-120}$.

c. $ {M_\gamma\over M }=0$.

d. $ {\alpha}\sim 10^{-2}$.

e. $ {M_e\over M_p}\sim 10^{-3}$.

The values of these parameters   are mostly mysterious. Nonetheless,  the
existence of these small numbers is essential for our understanding  of the
physical universe. The small value of ordinary masses to the Planck mass, the
\lq\lq hierarchy problem",
 is the reason that  gravity is such a weak force  weak at low
energy.  The cosmological constant is probably zero, certainly incredibly
small--but  we know not why.  Similarly the small ratio of the electron mass to
the proton mass, or the value of the fine structure constant have no
explanation. The only one of the small parameters we do understand is the
vanishing photon
mass, a consequence of an unbroken gauge symmetry.

If it was not for the existence of these small
parameters we  would have had a lot of
difficulty in developing  and understanding
physics. We would also not  be here.  For example, because of (a)
gravity is very  weak and we can ignore it at low energies.
This is good because it is a complicated force.
Due to the large hierarchy  strings look like particles at low energy, so that
we have had the luxury of first understanding   simpler   particle theories
before having  to
face up to
understanding string theory.
Another consequence of the
hierarchy is
that   big stars,
planets and macromolecules can exist. If the Planck mass and the  proton  mass
were comparable
then stars containing more than a few nucleons would collapse to black holes.

Because of (b), the smallness of the cosmological constant the
universe is so big.  If $\Lambda$ had its natural size the
the universe would be tiny and  we
would not be here. Because of(c), (d) and (e)    the
macroscopic objects exist and in the macroscopic world, to a good
approximation,    classical  non-relativistic physics is a good approximation.
Much of the development
of physics relies on a series of approximations
and hierarchies that follow from the existence of
these small numbers.

The same thing is true in QCD, a beautiful but complicated theory.
All of the progress that  we
have  made, as well  the progress that we
envisage in understanding  QCD  is based on the existence of small parameters.
What are
the small parameters in QCD?  After all QCD is a
theory that contains  no  adjustable parameters, except for the masses of
the quarks which are not truly important to the dynamics of the theory.  The
gauge coupling is not  a free parameter; it
runs with energy.  Nonetheless, there do exist many small parameters in QCD,
whose existence has helped us considerably.

The first set of  small parameters is are the ratio of the  light
quark masses relative  to the QCD mass scale: ${m_u , m_d/  \Lambda} \sim 1/20$
{}.
We might even include the strange quark mass in this list.
It is because these masses are so small
 that   chiral $SU(3)\times SU(3) $ is an approximate
symmetry of the strong interactions.  $SU(3)\times SU(3)$ is not a fundamental
symmetry of QCD; rather it is    an accidental
consequence of the small quark masses. Nonetheless it played
a crucial historical  role
in discovering quarks and understanding the
structure of the strong interactions leading to QCD.
It continues today to  provides the basis for an extremely
useful tool---chiral perturbation theory.

The second small parameter is the strong coupling constant $\alpha_s(Q^2)$ at
large momentum transfers, which, due to asymptotic  freedom, is  small .
This small parameter is even more useful than
the  others since it is adjustable.  One can always
make it smaller by  increasing the  energy, although it
only decreases logarithmically.  The fact that this
parameter is $\sim 1/10$ at high energies means
that we have the ability to do calculations with the aid of ordinary
perturbation theory. Much evidence of the agreement of these
calculations with
experiment was  presented in this meeting.

         Recently it has been understood that there is
another small mass parameter, which is very useful to
exploit, namely the  ratio of
the QCD mass scale to a heavy quark mass, such as  the
charm quark or the bottom quark.  This small parameter is of
 order of $\sim 1/5 \, \, {\rm to} \, \, 1/15 $.
 When this parameter is small,  {\it i.e.}
when  $N_Q$   quark masses are much larger than the QCD mass scale,
then QCD acquires an extra  $SU(2N_Q)$ symmetry.
This is a very beautiful recent development, only three
years old,  whose applications to the
enormous amount of emerging data on heavy quark
spectrum and decay amplitudes have recently exploded. The basic
observation is that in the limit where   the heavy quark
has an infinite mass compared to the
QCD mass scale,  then when one considers properties of
hadrons consisting of some number of heavy quarks,
one  can neglect the momentum of the light quarks
compared to the mass of the heavy quarks.
This can be summarized by an effective theory, called the heavy quark effective
theory,  which has an enhanced symmetry-- an
$SU(2N_Q)$  symmetry group which rotates the flavor
and spin degrees of freedom of the heavy quark.
The heavy quark, being so massive,
just sits in the hadrons  as a static color source; none of the physics
depends on its flavor or on its spin--thus the symmetry.
We heard in this meeting of the
very beautiful and successful applications of this
symmetry to discussing, cataloging and analyzing
the spectrum of heavy quarks and their  leptonic and semi-leptonic decay
amplitudes.   When   combined with chiral symmetry,  useful in analysing the
decay amplitudes of B mesons   to states containing pions
which are the Goldstone bosons
  of the chiral symmetry, you get very powerful
tools. This is a marvelously beautiful subject
that is developing rapidly. Unfortunately, like most
symmetries, it does not tell you much more more about
the theory than that it contains this   symmetry.  Nonetheless, it is
something that the theorists who developed this
should be very proud of, a marvelous way of
confronting the theory with the enormous amount of
data that is now  coming out.

        Another small parameter that has not been fully
explored   within QCD is  the Bjorken scaling variable $x$,
or rather  $1/ \ln(1/x)$.
 When one measures deep inelastic scattering   in  the region of small
x, one is probing a region  not dissimilar
 from the region of diffraction scattering, of Regge behavior.
This is the region  where another
small parameter, namely the ratio of the QCD mass
scale to the energy together with ratio of    the momentum transfer
 to the energy are small.  This region is now  opening up,  partly due to the
new
tool provided by   Hera.  Hera is a marvelous
tool for exploring large $Q^2$ physics and testing
quantum chromodynamics to very high precision. But it is also
  a marvelous tool for exploring small $x$ physics
where new phenomena might appear.  It is
amusing  that it might very well be that
the  interesting physics that will  emerge from  Hera will be
the small $x$ physics, unlike the first experiments in
deep-inelastic scattering, which revolutionized our view of  the  strong
interactions by  discovering  scaling at large
$Q^2$.

At the moment Hera  has
too  low a  luminosity to accumulate enough   events for
precision tests of QCD. However, at small $x$,
where  most of the cross-section is, and for
reasonably large values of $Q^2$, they have already
observed a quite
remarkable phenomenon. They have seen  a large number of events
with a large rapidity gap,  a phenomenon which has also been seen in
$D_0$ and in UA8.  Both groups at HERA have reported that of the order of
5\% of the events with small $x$ and large $Q^2$
have a large rapidity gap between fragments of
the virtual photon and  the fragments of the
proton.  These events look very different than the standard deep
inelastic scattering events that we understand in
perturbative QCD. Such   events  look very much
like diffraction scattering, indeed the
energy dependence is constant.
What is perhaps surprising is that these events scale
({\it i.e.} show no power falloff in $Q^2$.)
This phenomenon is sometimes  associated with the
hard structure of the Pomeron.  The Pomeron,
responsible for diffraction scattering  for
fixed $Q^2$, is still one of the most
mysterious and least understood aspects of strong
interaction physics. There have been many
theoretical ideas, mostly from the Leningrad school,
which lead one to expect such a phenomenon.
I regard this as a very interesting development.
The  new experimental tool provided by HERA
should
motivate theorists to once again attack the problem of
the Pomeron within QCD  and perhaps we
might finally get an  handle on diffraction scattering.

Another small parameter in QCD  is simply
$ 1/N_c= $ 1/the number of colors. This is perhaps
the nicest of all the  small parameters because its value is
independent   of energy and
can be used as an expansion parameter for both short and long range dynamics.
Since there are   three colors the actual
expansion parameter is $(1/N_c)^2\sim 1/10$.
It is well known   that in that in the limit where this
parameter is regarded  as  very small QCD has
same qualitative features as it does for $N_c=3$, but the calculation of
these features becomes much simpler.  In fact, there
are many $1/N_c$ theorems or relations in QCD that work very well.
The $1/N_c$ expansion is perhaps the only hope we have   that might
 allow us total analytical control
over the theory.
Recently  I have been trying to construct, in this $1/N_c$
expansion, a string picture of QCD. That one might hope to
describe the infra-red confining dynamics of QCD by means
of an effective string theory  should not
be too much of a surprise. This is
an old idea; in fact string theory originated from pion
nucleon phenomenology.  This is not the fundamental
string which contains quarks and gluons, rather  a
string which describes hadronic flux tubes  and which
exhibits the nearly linear Regge trajectories.
Such a picture is both suggested by hadronic phenomenology
and by   the  $1/N_c$ expansion of QCD.
The hadronic string theory (which I call SCD=String Chromodynamics) would
be quite different the fundamental string theory,
it would not contain gravitons, gauge mesons or dilatons.
 My strategy to  find the formulation of SCD has
far has been to work  in two
dimensions.   Two dimensional QCD much simpler than the real world,
but it has many of the same features---confinement, an infinite spectrum
of hadrons,etc.
The idea is to develop the string theory in two dimensions if possible,
It turns
out that in two dimensions you can  indeed
rewrite the theory as a string theory.  The plan is
then to understand two dimensional SCD well enough to continue to
four dimensions.  These results are very encouraging but this  program is just
beginning.

        There are other small parameters in QCD that arise
if you   push the theory to extreme
conditions.  For example the ratio  of the QCD mass scale to the
temperature or to the baryon density. If you could heat up quark matter,
so that this ratio would be very small, then the vacuum
will  undergo a deconfining phase transition as well as a
chiral symmetry restoring phase transition. At
very high temperatures the vacuum should be understandable as
a quark gluon  plasma. A similar  transition should take at high baryon
density.
The  precise dynamics  in regions of   extreme
conditions is quite complicated and not that well
understood.  Here there are potential  applications here to heavy
ion physics and dense nuclear physics.

Finally there is another small parameter that
is exploited all the time, namely  is the  cost/size of modern
computers.  This is the parameter that  lattice
calculations depend on.   It is time dependent,
decreasing steadily.

\section*{Beyond the Standard Model}

Let us  now  go  beyond the standard model.   I shall
 discuss   very briefly supersymmetry and
string theory.  Supersymmetry is rapidly coming into
vogue. It is a little scary how rapidly  many
 experimentalists have adopted  supersymmetry as  the standard
non-standard theory.  Some people think that is a bad development,
since it might mean that they are being hoodwinked by the theorists and
  are closing their eyes to other
possibilities that the theorists have not thought  of.
I think not.
Supersymmetry provides us with  definite and  relatively precise
predictions of almost impossible to measure
processes. It therefore poses a  real challenge to
experimentalists when they design new
detectors. The alternative, given the constraints of reality,
is not to prepare for all possibilities but for none.

 Why do the theorists love
supersymmetry? There are at least three reasons. First it is a
beautiful extension of ordinary Poincar{\'e} symmetry that
offers the only known explanation  why fermionic
matter exists. It makes the existence of fermions as much a consequence
of a deep  symmetry as the existence of gauge bosons. Second,  it is
an automatic and necessary component of string
theory.  In string theory supersymmetry does not just
explain the existence of photons; it is required for
the existence of fermions.  If you want to construct
a string theory that contains fermions, it has to be
supersymmetric. Finally, most importantly from
a phenomenological point of view  it offers
the most plausible solution of the hierarchy problem,
the enormous ratio of the GUT  scale to  the low energy scale.
In this attractive scenario  the masslessness of the
Higgs boson is protected from developing at the
unification scale by supersymmetry, whose
breaking then drives the electroweak  symmetry breaking at a
much lower energy. Consequently, this mechanism  requires  that supersymmetry
be
observable at low energies. The superscale, that determines the
mass splittings between particles and their
supersymmetric partners must be of order 100 Gev.

 The exciting
possibility therefore exists  that the discovery of
supersymmetry is just around the
corner, at the
next generation of hadron colliders.
There is already increasing  evidence for supersymmetry.
First,   the high-precision measurements of the standard
model parameters can be used to   extrapolate standard model couplings
to arbitrarily to very high energy.  These extrapolations
certainly contradict the simple predictions of
minimal GUT's without supersymmetry ($SU(5)$). On
the other hand they are in beautiful agreement with the
simplest predictions of minimal supersymmetric
unification.  In this supersymmetric $SU(5)$ model, when the couplings are
extrapolated by 12, 13 orders of magnitude, they  all come
together at a scale of $2 10^{16}$ Gev,  very close to the Planck scale with
a coupling that is still reasonably weak ($ g^2/4\pi \sim 1/26 $.)
The increase in the GUT scale is
due to the  screening of the strong coupling by the
extra supersymmetric matter particles,
therefore pushing its unification with the
electroweak coupling to higher energy.  This agreement is very
nice. Of course it cannot be taken as proof or even as very
strong evidence for supersymmetry, but if it  had not
worked,  even to the extent that standard $SU(5)$ does not work, it
would be regarded as strong  evidence against supersymmetry.

 This unification of couplings is also
nice in the context of string theory,   which
is automatically supersymmetric and whose energy scale is
usually identified with the Planck mass scale.
You might think that there is a disparity,  since
the unification scale would appear to be quite a bit below the
Planck mass scale, which is normally quoted to be
$10^{19}$Gev.   However, in the case of the heterotic string theory,
the relation between the Planck
mass scale $M_{\rm Planck}$  and the string scale, $M_{\rm string}$,
is dependent on the value of the
coupling: $M_{\rm string}= 2/\sqrt{\alpha'}=gM_{\rm Planck}/\sqrt{8\pi}$.
In addition there exist \lq\lq  threshold corrections",  slight shifts in the
scale in which the couplings are unified because of
the effects that come from  integrating out
other heavy particles of the theory to one loop order before you make a
comparison with   the effective low energy field
theory. The net result is that one
finds that the string unification scale is $\sim 3 10^{17}$ Gev,
only   a factor of 20 from what
is predicted by this 13 to 14 orders of magnitude of
extrapolation.  I regard this agreement as  a great
success and a  great comfort to string theorists.

Another very nice development that makes
supersymmetry look better has to do with the top quark.
It is not just that the standard
model parameters, when more precisely measured,
agree with supersymmetric GUT, but also   the fact that the
top quark is so heavy makes supersymmetry look much better.
One reason is that with a heavy top there is a natural mechanism for producing
the electroweak scale of $10^2$ Gev from the GUT scale of
$10^{16}$ Gev.  This is an old argument,  which showed
that, in a theory with a heavy top quark, the radiative
 one loop corrections, arising from the top quark
loop,  to the Higgs mass
matrix, $\Delta m^2_{\rm Higgs} \sim -m_t^2 \ln(m_t/\mu)$
(which are negative  because  of the fermion  loop), can become large enough
 to drive the Higgs potential
unstable and produce a Higgs mass at the electroweak
scale.  This is  a very simple,
beautiful  and natural   explanation of the hierarchy, that
becomes credible once the top quark
mass is as heavy as it seems to be.

Of course, there are
problems with supersymmetry. Most important  we do not understand
how it is broken, certainly not in the context of string theory.
Another problem is that
supersymmetric models tend to produce dangerous flavor changing processes.
At this conference there was reported the detection
of the radiative decay of the bottom quark
to the strange quark: $b \to s \,\, \gamma$,
which is a threat to
supersymmetry since  supersymmetric models contain   a charged
Higgs which can contribute to this decay.
The situation is that CLEO has placed a limit ($<5.4 10^{-4}$)
on the inclusive branching ratio of the $b \to s \,\, \gamma$ decays,
and has observed the decay $B\to K^* \,\, \gamma$
with a branching ratio of $ (4.5 \pm 1.5 \pm.09 ) 10^{-5}$.
This experiment  places a limit of the masses of the charged
Higgs ($m_{H^-} >250$Gev),  which seems too  high for standard SUSY models.
I have been told that this problem is easy to circumvent,
so it is probably not a  a worry.  It is interesting  that this case
 stands out because supersymmetry, unlike other
nonstandard models, has so far managed to
squeak through.  At a much   more fundamental level, it is
very unfortunate that supersymmetry did not provide
an answer for the vanishing of the cosmological
constant, but that is not really an argument against it.

So, in my opinion, the most exciting prospect for the
SSC-LHC is the discovery of supersymmetry.
Somehow we have been selling these machines to the
public in the wrong way. The SSC is not only the machine whose
purpose is to discover the Higgs boson but also the instrument that  could
discover supersymmetry.  The United States
is going to spend a lot of money on exploring the
ordinary three  dimensions of space,  but
at the SSC we will  discover not just more of ordinary
space but new dimensions of space.  Of course, the new dimensions will
be   fermionic but they are just as real as the ordinary bosonic  dimensions of
space and time.  If we can only convey that feeling of exploration to the
public, that we are enlarging space and time, maybe
we can convince them to support the SSC.

\section*{STRING THEORY}

Turning now  to string theory I flash my usual list of the
achievements of string theory:

$\bullet$ String Theory   provides a consistent, logical and rich  extension of
the
conceptual structure of physics.

$\bullet$ String Theory   provides  a consistent and  finite
theory of quantum gravity.

$\bullet$ String Theory   provides a  rich structure which might yield
a unique and  comprehensive description of the
real world.

Let me discuss the second achievement first.
The existence of a finite, consistent theory of quantum gravity   within
string theory is, at  the very least,    an existence proof
that gravity and quantum mechanics are mutually consistent. Many
people suspected that this was not the case,
so it is important to have such an example.
In addition, stringy gravity is a
very interesting theoretical tool.   It
has motivated the construction of many  low energy models of
(super)gravity with additional fields  such as the dilaton and axion.
It has also provided
 some   soluble toy models of two  dimensional gravity.
These provide
 a very useful  theoretical laboratory for discussing interesting
and important conceptual issues of quantum gravity,
especially those related to the fate of black holes.

        As for phenomenology, string theory provides
interesting unification scenarios   that are often quite
different from those emerging from ordinary field theoretic GUT's.    I
think the most interesting thing to focus on is not the
detailed prediction of numbers,  because you can not
trust these, but rather those aspects of nonstandard
physics that are nonstandard.  These include the existence
of discrete symmetries, many possibilities of flavor changing processes,
new forms of CP violation, the existence of extra matter: Z's and gauge
singlets
and the possibility of strange solitons.

We also heard about the very amusing development in which
 string theory turns out to  be a useful way of viewing field
theory. After all one can always view some field theories
as the low energy limit of string theory. Remarkably,
when one looks at field theory in this light one
finds that the perturbative calculational
devices of string theory   organize the field theoretic
calculations in a very simple
way. This has been  fruitfully used to develop improved
calculational tools
for perturbative QCD.

But I want to come back to the revolutionary aspects
of string theory,
since I am supposed to talk about {\em  vision}. There
have been two major revolutions completed in this century: relativity,
special and general,
and quantum mechanics. These were associated with two of the three
  dimensional parameters of physics: $\hbar$,
Planck's quantum of action and  $c$,
the velocity of light. Both involved
major conceptual changes in the framework of physics,
but reduced to  classical non-relativistic physics
when $\hbar$ or $1/c$ could be regarded as small.
The final dimensional parameter is Newton's gravitational constant, that sets
the
fundamental (Planck) scale of length or energy.
Some of us  believe that string theory is the
 revolutions associated with this  last of the dimensional parameters of
nature.
In eventually understanding string theory we
will have to undergo a discontinuous conceptual
change in the way we look at the world
similar to that that occurred in the development
of relativity and quantum mechanics.

You might regard this, if true,  as very bad news. The evolution of
 quantum mechanics,  from Planck, Einstein and Bohr (1898-1913) to the final
theory
(1926), required  an enormous amount of
experimental input.   It is very hard to imagine,
even with  the SSC and much luck,  an enormous amount of
experimental input  that will provide us with direct information regarding
Planck mass
physics.  However,
we should also remember that the other revolution  of this century, that having
 to do with relativity---special and
general---did not require an enormous amount of
experimental input.  In fact, that conceptual revolution  occurred
after one already had a theory,  Maxwell's theory of  electromagnetism (1865),
which was a relativistic theory.  Einstein only had to understand it
better (1905).  One can argue that string theory lies
somewhere in between these two examples.  Every historical revolution  (as we
have witnessed so dramatically in recent years) is different.
The same is true in  physics.  The development
of string theory might a mixture of the quantum and relativistic revolutions.
We already have a handle on string theory, the
perturbative  definition and  construction of the theory.
But we do not yet have the full fledged formulation of the theory at all.
The construction of the ultimate formulation of string theory
might require a discontinuous conceptual jump, but
we might be able to make this jump just as Einstein could develop relativity
 from electromagnetism without much
experimental input.  As you know,  the
Michelson-Morley experiment, although of fundamental importance, did not
 influence Einstein  very much. What did  influence
him were Maxwell's equations.

There are many indications that such conceptual
jumps will be required in the theory. Most likely we will be forced to
change our view of the structure    of space-time.  One indication
that I am particularly fond of  is the stringy analogue of the Heisenberg
microscope for probing the structure of particles, or the structure of
space-time itself.  At low energies the
accuracy of measurements of position is
dominated by quantum mechanics, $\Delta x \sim \hbar /p$,
but at high energies the accuracy
is dominated  by the fact that strings themselves are
extended objects.  At high energies  they smear out,
and the Heisenberg uncertainty principle is replaced by,
\begin{equation}
\Delta x  \approx {\hbar c \over E}  + {G_{\rm Newton} E \over g^2 c^4} .
\end{equation}
  Thus  the accuracy of position measurements  at high energy is limited
by  a stringy smearing term
which is not quantum mechanical.
Consequently one can not  measure  distances smaller
than (approximately) the Planck length.

Other indications of the nature of the conceptual revolution
hinted at in string theory
abound. One is  {\em duality}.  If
strings are compactified on a circle  of radius $ R$
or on a circle of radius $l_P^2 / R$
one finds identical physics, again suggesting that
distances smaller than $l_P$ make no sense.
This symmetry is very stringy, relating ordinary
excitations of the string to solitonic states arising
from the winding of the string about the compactified dimensions.
Another indication arises from these of the very
beautiful mirror symmetry,  discussed by Brian Greene,
a very stringy symmetry which equates the physical
content of  apparently different compactified
spaces. This symmetry  has been used for very mundane things, such as
 calculating Yukawa couplings and determining fermion
masses in  superstring models.  It has also been used
to demonstrate the string physics goes smoothly through a change of  topology
of
space.  This is is almost a proof  that
there is no way one can understand string theory in
its full structure within the standard conceptual framework we
have of space and time.  Another
indication arises  from the size of nonperturbative effects
in string dynamics.  There is a connection between the high order behavior of
perturbation theory and the nature of the
nonperturbative dynamics.  In string theory we  unfortunately know
nothing  about nonperturbative
dynamics. That is the main reason we can not
compare the theory with experiment. But we can construct  the perturbation
theory. One  finds a big
difference between the large order behavior of field theory and of string
theory.
String  perturbation theory is  of twice as divergent as field theory, which
  translates   into the fact that the
nonperturbative phenomena associated with these
divergencies of perturbation theory must be different in nature.
The magnitude of the non-perturbative effects will be much stronger
for small coupling $g$,  $ \exp(-1/g) $ as opposed to  $ \exp(-1/g^2) $.
Such effects
cannot be described by ordinary field theory, non-perturbative
string theory will be very different
from ordinary field theory.

\section*{ PREDICTIONS}

I shall  end this talk with a set of predictions. Since this
conference is thirty years old  I thought I might
predict thirty years into the future---but that is much
too far.  So I will  predict halfway, say  fifteen years into the
future,  to the year 2008. First, ten experimental predictions, in order
of likelihood:

1. The top quark    will be old news and will have a
mass of $160\pm 20$ Gev.

2. $  \epsilon'/ \epsilon$ will finally be measured and will not be zero.
In addition the three B-factories will have found new  manifestations of
CP violations.

3. At least 2 light Higgs particles will be found after a few years of running
at the SSC.

4. There will be convincing evidence for the existence of
supersymmetric particles.

5.  The astrophysicists will
finally have determined that $\Omega = 1$ with an
accuracy of 10\%.  Particle physicists will
understand that this mass density is composed of   a combination of baryons,
axions, neutrinos and neutralinos with various weights (some of which
could be zero.)

 6.  The observation of neutrino
oscillations will verify the MSW mechanism and be consistent with the solar
neutrino problem.

 7.  There will be evidence of the
quark  gluon plasma and of a chiral phase transitions in
  heavy ion collisions.

8.  Some number of  new Z-mesons    will be  discovered.

9. There will be cloudy evidence of
superstrings.

10.  Finally, the most surprising and
strange of   all  predictions---there will be a
real surprise!

\bigskip
Ten more predictions for theory:

1.   Lattice  gauge theory, armed with Mega-Tera-flop computers,
will  calculate the hadronic
spectrum from QCD with  1\% precision.

2. Analytic  treatments in  QCD will developed to describe  small $x$ physics,
Regge behavior and hadronic  fragmentation functions.

3. A
nonperturbative treatment of QCD based on the $1/N_c$
expansion and/or stringy QCD will be developed.

4.  There will exist a quantitative
understanding of  the cosmological origin of baryons.

5.  There will exist a quantitative
understanding of  the cosmological origin of
density fluctuations leading to the large scale structure of the universe.

6.  String field theory will  begin to be a useful tool and will
illuminate the underlying symmetries of  the theory.

7. New mechanisms of string supersymmetry breaking will be discovered
leading to  new and
definitive low energy models.

8.  The  conceptual revolution arising from the
nonperturbative formulation of string theory will be in full swing,
revolutionizing the concepts of space-time geometry.

9.  The fate of evaporating black holes will be understood without modifing
the
basic principles of quantum mechanics.

10.  And, most unlikely, we will understand why the cosmological constant is
zero.
\section*{Conclusions}

  I often conclude   talks like this one with  a discussion
of whether we should be  optimistic or pessimistic regarding the
development of our field.    I
note that physicists tend to  oscillate between extreme optimism
and extreme pessimism.
They  are optimistic when they have
discovered new phenomena or a new theoretical technique and are sure that all
problems will be cleared up in short order.  They are
pessimistic when they discover  problems with the
background or problems with the  theory and   conclude
that these problems are insuperable.  They  are very rarely moderate.
I usually describe  myself as
being  cautiously optimistic, although I often point out
that the optimist is one  who proclaims that we have the best of
all possible theories and the pessimist is one who  fears that this
is true.

        However, there is a large cloud hanging over this conference, blocking
out all the rays of optimism.
  I  will  conclude, therefore,
with a remark about the SSC.  I've been serving on
  the PAC committee at the SSC for the last
few years and I have had  the opportunity to observe
the heroic efforts of our colleagues who are trying to
construct a new laboratory and to mount new
experiments at this facility. We must do
everything possible to help them in these very
difficult times.  But more than that, we must  do
everything possible to help ourselves.  If the
SSC fails, as it might,  there would be a tidal
that could sweep away our field.  It is very
hard to think of strategies for the long term survival
of particle physics as an experimental or theoretical
field without   the SSC.

Since I do not want to end
on  such a pessimistic note, I will end with David
Hilbert's epitaph:

\smallskip
\centerline {\bf We Must Know--- We Shall Know }
\eject

\section*{QUESTIONS}

\smallskip
Q:  I just want to try to understand the modification
of the uncertainty principle  in string physics.  Do you expect this
to be changed with the new nonperturbative
formulation that you are predicting,  or is it pretty
much a fixed prediction at the current state of the
art?

\smallskip
A: The argument I gave  involved much hand waving, much as
 Heisenberg's arguments were  hand waving.  Now,  within quantum mechanics you
can
formulate  quite precisely what you mean by
the uncertainty principle.  What I actually
believe is that in the new string theory that will eventually arise, much
like quantum mechanics replacing the Bohr theory,
space and time will not be primary constants.  Space and time
will emerge as concepts one can use you use to describe physics
under certain circumstances.  They will not be primary concepts.
But  I have absolutely no idea what a precise
formulation of the deviations from ordinary geometry will be provided by string
physics.

Q:   I'd like to ask if there is any prospect of raising
money from Congress by selling space on the
compactified dimensions?

A: Perhaps. Another suggestion would be to put some of these  congressmen, and
I have a little list,  into those compactified dimensions.

\end{document}